\documentclass[twocolumn,aps,prc,preprintnumbers,showpacs,showkeys,nofootinbib,superscriptaddress,fleqn,floatfix,tightenlines, 10pt]{revtex4-1}
\usepackage{amsmath,amsfonts,amssymb,amscd,amsxtra,amsthm}
\usepackage{natbib}
\usepackage{graphicx,subfigure}  % Include figure files
\usepackage{epstopdf}
\usepackage{dcolumn}  % Align table columns on decimal point
\usepackage{bm}          % bold math
\usepackage{slashed}
\usepackage[utf8]{inputenc}
\usepackage{CJK}
\usepackage{mathptmx}
\usepackage{mathrsfs}
\usepackage[normalem]{ulem} % \sout{old text} for strikeout
\usepackage[dvipsnames]{xcolor} % For blue in-text comments and
\usepackage{array}
\usepackage{multirow}

\begin{document}
%\preprint{Arxive}
\title{Three-body resonant states of $^9_{\Lambda}$Be
with $\alpha +\alpha +\Lambda$ cluster model}

\author{Qian Wu}
\email[E-mail: ]{wuqian@smail.nju.edu.cn}
\affiliation{Department of Physics, Nanjing University, Nanjing
  210093, China}

\author{Yasuro Funaki}
\email[E-mail: ]{yasuro@kanto-gakuin.ac.jp}
\affiliation{College of Science and Engineering, Kanto Gakuin University,
Yokohama 236-8501, Japan}
\affiliation{RIKEN Nishina Center,RIKEN,2-1 Hirosawa,351-0115
  Saitama,Japan}

\author{Emiko Hiyama}
\email[E-mail: ]{hiyama@riken.jp}
\affiliation{RIKEN Nishina Center,RIKEN,2-1 Hirosawa,351-0115
  Saitama,Japan}
\affiliation{Department of Physics,Kyushu
  University,819-0395,Fukuoka,Japan}

\author{Hongshi Zong}
\email[E-mail: ]{zonghs@nju.edu.cn}
\affiliation{Department of Physics, Nanjing University, Nanjing
  210093, China}
\affiliation{Joint Center for Particle, Nuclear Physics and Cosmology,
  Nanjing 210093, China}
\affiliation{State Key Laboratory of Theoretical Physics, Institute of
  Theoretical Physics, CAS, Beijing, 100190, China}
%\maketitle

\begin{abstract}
We study the structure of $^9_\Lambda$Be in the framework of three
body $\alpha+\alpha+\Lambda$ cluster model using YNG-NF interaction
with the Gaussian expansion method. Employing the complex scaling method,
we obtain the energies of bound states as well as energies and decay widths of the resonant states.
By analyzing our wave functions of bound states and resonant states,
we confirm three analogue states of $^9_\Lambda$Be pointed out by Band${\rm \bar{o}}$ and Motoba {\it et al.}
\cite{motoba1983,motoba1985,bando1983}, $^8$Be analogue states,
$^9_{\Lambda}$Be genuine states and $^9$Be analogue states.
The new states of $^9_\Lambda$Be are also obtained at a high energy region with
broader decay widths.
\end{abstract}
\maketitle
\section{Introduction}

One of the main goal in hypernuclear physics is to explore the new dynamical feature
by  an addition of
a $\Lambda$ particle.
Since there is no Pauli  principle between nucleons and a
$\Lambda$ particle, participation of $\Lambda$ particle in nuclei
gives rise to more bound states. As a result,
we have significant contraction of nuclear cores.
We call this phenomena as `glue-like' role of a $\Lambda$
particle.
Such a study for the energy stability  in light hypernuclei has been studied
in Refs.\cite{motoba1983,bando1983,motoba1985,hiyama1995,hiyamanpa2012} for
stable nucleus plus a $\Lambda$, and for neutron-rich nuclei plus a
$\Lambda$. One of typical example is a combination of
$^6$He and $^7_{\Lambda}$He.
The core nucleus, $^6$He is known to be a halo nucleus whose observed
binding energy
of the ground state is $-0.96$ MeV, weakly binding with respect to
the $\alpha +n+n$ breakup threshold.

In Ref. \cite{hiyama1995}, one of present author (E. H.) predicted
that the ground state of $^7_{\Lambda}$He should become
more bound due to the glue-like role of a $\Lambda$ particle
and that the $\Lambda$-separation energy,
$B_{\Lambda}$ was 5.44 MeV within the framework of
$^5_{\Lambda}{\rm He}+N+N$ three-body model (Afterwards,
within the framework of $\alpha +\Lambda +N+N$ four-body model,
she predicted $B_{\Lambda}= 5.36$ MeV\cite{hiyama2009}.)
In 2013,  they observed this neutron-rich $\Lambda$ hypernucleus
 for the first time
at JLAB by $^7{\rm Li}(e,e'K^+) ^7_{\Lambda}{\rm He}$ reaction and
reported the observed $B_{\Lambda}=5.58 \pm 0.03$ MeV \cite{JLAB2013},
which was found energy gain to be 5.58 MeV due to the participation of
a $\Lambda$ particle.

Also, dynamical contraction by the addition of a $\Lambda$ particle has
been studied by many authors \cite{motoba1983,bando1983,motoba1985,hiyama1999prc,hiyama1995,lubingnan,isaka,funaki-PLB}.
Historically, in Refs.\cite{motoba1983,motoba1985}, they
studied light $p$-shell $\Lambda$ hypernuclei such as
$^7_{\Lambda}$Li, $^8_{\Lambda}$Li, $^8_{\Lambda}$Be and
$^9_{\Lambda}$Be within the microscopic $\alpha +x+\Lambda$
three-body cluster model ($x=d,t,^3{\rm He}$)
together with the $\alpha +x$ two-body cluster model for the
nuclear core.
They pointed out that reduction of the $B(E2)$ strength led to
the contraction of the hypernuclear size since the $B(E2)$ was
proportional to the fourth power of the distance between the
$\alpha$ and $x$ clusters and then they predicted that $\alpha$-$x$ distance
in above $A=7$ to 9 $\Lambda$ hypernuclei should be reduced by about 20 \%.
Afterwards, one of present authors (E. H.) proposed to experimentalists to
measure $B(E2)$ of $5/2^+ \rightarrow 1/2^+$ in $^7_{\Lambda}$Li and
predicted $B(E2)=2.42 e^2{\rm fm}^4$ within
$^5_{\Lambda}{\rm He}+N+N$ three-body model.
At KEK, measurement of this hypernucleus was done successfully
and they reported $B(E2)= 3.6\pm 0.5e^2 {\rm fm}^4$,
 which confirmed the shrinkage effect by an addition of the $\Lambda$ particle \cite{li7exp}.

Another interesting issue to study dynamical structure is to find new states
due to the injection of  $\Lambda$ particle, which does not exist in normal nuclei.
For this study, Band${\rm \bar{o}}$ and Motoba et al. \cite{motoba1983,bando1983,motoba1985} investigated the level structure of
$^9_{\Lambda}$Be within an $\alpha +\alpha +\Lambda$ three-body model
and categorized three-types
of states, ` $^8$Be analogue states',  $^9$ Be analogue states', and `genuine states', according to the SU(3) shell model classification. The $^8$Be analogue state corresponds to the SU(3) irreducible representation with $[s^5p^4](\lambda, \mu)=(4,0)$, where the $\Lambda$ particle occupies the $(0s)$-orbit. Band${\rm \bar{o}}$ and Motoba {\it et al.} defined two $^8{\rm Be}+\Lambda(0p)$ configurations as `$^9$Be analogue states' and `genuine states', corresponding to the [$s^4p^5$]($\lambda, \mu)=(3,1)$ and [$s^4p^5$]($\lambda\mu)=(5,0)$ irreducible representations, respectively. The latter is a new symmetry called `super-symmetric' by Dalitz and Gal \cite{daliz-a,daliz-b}.
In view of weak coupling, or of nuclear clustering, in these configurations the $\Lambda$ particle is supposed to orbit around a well developed $\alpha$-$\alpha$ core since the $^8{\rm Be}$ core has a well developed $2\alpha$ cluster structure. In particular, in the $^9{\rm Be}$-analogue and genuine states, the $\Lambda$ particle occupies two kinds of $p$-orbit, perpendicular and parallel to the deformation axis of $\alpha-\alpha$ core, respectively.

Here it should be noted that
Band${\rm \bar{o}}$ and Motoba  {\it et al.} \cite{motoba1983,bando1983,motoba1985} obtained
three categorized states, almost all states of whose are resonant states
above the lowest threshold $^5_{\Lambda}{\rm He}+\alpha$
using bound state approximation with restrict configuration,
that is, they took only one Jacobian coordinate of $(\alpha \alpha)-\Lambda$
channel (See in Fig.1) using small number of basis functions for
$\alpha$-$\alpha$ (${\bf r_2}$ of Fig.1)
and $(\alpha \alpha)-\Lambda$ (${\bf R_2}$ of Fig.1) coordinates.
It is considered that bound state approximation works for narrower resonant states,
but it would be difficult to obtain broader resonant states.

To obtain resonant states theoretically, there have been many achievements.
A successful method is Complex scaling method (CSM)\cite{CSM1,CSM2,CSM3,CSM4,CSM5,CSM5} .

For this purpose, recently, using CSM and Gaussian Expansion Method (GEM),
we obtained all of possible resonant states of
$^9_{\Lambda}$Be within the framework of $\alpha +\alpha +\Lambda$
three-body model \cite{few-body2019}.
In Ref.\cite{few-body2019}, we found
energies and ordering of some resonant states were consistent with
those by Motoba {\it et al.}, and some were different from results by Band${\rm \bar{o}}$ and Motoba {\it et al.} \cite{motoba1983,bando1983,motoba1985}.
However, we have not analyzed
three- types of category for states  obtained in Ref.\cite{few-body2019}
in detail.

Therefore, in this work,
we calculate energy spectra of bound states as well as of resonant states
with the $\alpha  +\alpha +\Lambda$ three-body model $+$ CSM using
the same $\alpha \Lambda$ interaction as used in Ref.\cite{few-body2019}.
By comparing our wave functions with those of the SU(3) shell-model wave functions, we confirm that the $^8{\rm Be}$ analogue states',  $^9{\rm Be}$ analogue states', and `genuine states' appear, as discussed by Band${\rm \bar{o}}$ and Motoba {\it et al.} \cite{motoba1983,bando1983,motoba1985}. We also find new states by $10 \sim 20$ MeV above the $\alpha +\alpha +\Lambda$ three-body breakup threshold, which have never been pointed out by Band${\rm \bar{o}}$ and Motoba {\it et al.}

Finally, we discuss our calculated states in comparison with
observed data \cite{hashimoto-1998,Hhashimoto}.

This paper is organized as follows. In section \uppercase\expandafter{\romannumeral2}, we
introduce the realistic NN and $\Lambda$N interaction and the unique adjust of the some parameters.
After explaining Method employed,
 we show the results and the discussion of
$^9_\Lambda$Be.  Summary is given in Sec. IV.

\section{Method and Hamiltonian}
Since we investigate the $^9_\Lambda$Be in the framework of $\alpha+\alpha+\Lambda$ cluster model,
the hamiltonian then is defined as:
\begin{equation}
H=T+V_{\alpha_{1} \alpha_{2}}+\sum_{i=1}^{2} V_{\alpha_{i} \Lambda}+V_{\alpha_{1} \alpha_{2}}^{\text { Pauli }},
\end{equation}
where T is the kinetic energy operator. $V_{\alpha_{1} \alpha_{2}}$ and $V_{\alpha \Lambda}$ represent for
 the $\alpha$-$\alpha$ interaction and $\alpha$-$\Lambda$ interaction ,respectively. The $V_{\alpha_{1} \alpha_{2}}^{\text { Pauli }}$
 stands for the Pauli principle between 2$\alpha$s.

\begin{figure}[tb]
\begin{center}
\centerline{\includegraphics[width=180pt]{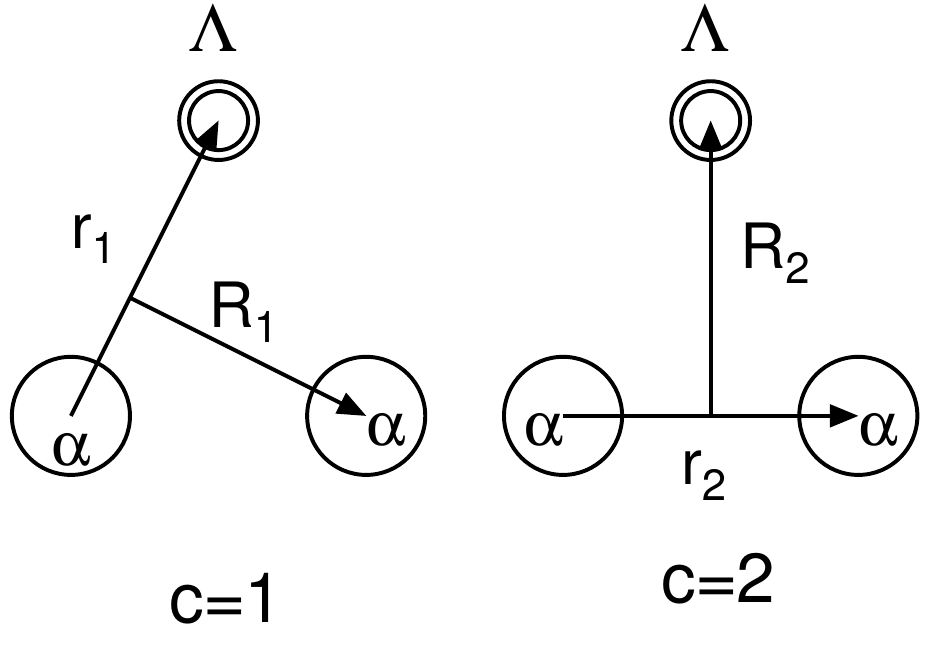}}
\end{center}
\caption{
Jacobian coordinates of $^9_{\Lambda}$Be with
 $\alpha +\alpha+\Lambda$ model
}
\label{fig:be9job}
\end{figure}

In order to solve the Schr$\ddot{\rm o}$dinger equation, the Gaussian Expansion method \cite{Hiyama2003GEM,hiyama1997ptp} enables us to use
two sets of Jacobian coordinates ($c=1\sim2$) of Fig.\ref{fig:be9job} in our total trial wave function:
\begin{eqnarray}\label{twf}
\Psi_{JM}=\sum^2_{c=1}\sum_I \sum_{\ell,L}\sum_{n,N}C^{(c)}_{n\ell NLI} {\cal S}_\alpha\{[\Phi^{(\alpha_1)}\Phi^{(\alpha_2)}] \nonumber \\
\times[\phi^{(c)}_{n\ell}({\bf r}_c)\psi^{(c)}_{NL}({\bf R}_c]_I\chi_{\frac{1}{2}}(\Lambda)\}_{JM},
\end{eqnarray}
where $\mathcal{S}_{\alpha}$ is the $\alpha$-$\alpha$ symmetrization operator and $\phi^{(\alpha)}$ is
the intrinsic wave function of $\alpha$ with $(0s)^4$ configuration. $\chi_{\frac{1}{2}}(\Lambda)$ is the spin
function of $\Lambda$. Since the energy splitting of $3/2^+-5/2^+$ is almost negligible measured by the
high-resolution $\gamma$-ray experiment \cite{akikawa2002,Tamura2005}, we neglect the spin-orbit force between
$\alpha$ and $\Lambda$ and simply regard $\emph{I}$ as $\emph{J}$. And the spatial part $\phi_{n\ell m}(\bf r)$ and $\psi_{NLM}(\bf R)$
have the form:
\begin{eqnarray}
&\phi_{n\ell m}({\bf r})=r^\ell e^{-(r/r_n)^2}Y_{\ell m}(\hat{\bf r}), \nonumber \\
&\psi_{NLM}({\bf R})=R^L e^{-(R/R_N)^2}Y_{LM}(\hat{\bf R}),
\end{eqnarray}
where the Gaussian variational parameters are chosen to have geometric progression:
\begin{eqnarray}
&r_n=r_{\rm min} a^{n-1},  \quad (n=1 \sim n_{\rm max}) \nonumber \\
&R_N=R_{\rm min} A^{N-1},  \quad (N=1 \sim N_{\rm max}).
\end{eqnarray}
Both the eigen energies and the coefficients $\emph{C}$ are obtained by using the Rayleigh-Ritz variational
method.

The $\Lambda\alpha$ interaction is obtained by folding the $\Lambda$N interaction into the $\alpha$ cluster
wave function. We use a so-called YNG interaction which simulated G-matrix $\Lambda$N interaction derived from
Nijmegen model f(NF)by the three-range Gaussian form as a function of $k_F$. The YNG interaction is given as:
\begin{eqnarray}
%\begin{split}
V_{\Lambda N}(r,k_F)=\sum\limits_{i=1}^{3}[(v_{0,even}^i+v_{\sigma\sigma,even}^i\sigma_{\Lambda}\cdot\sigma_N)\frac{1+P_r}{2} \nonumber \\
+(v_{0,odd}^i+v_{\sigma\sigma,odd}^i\sigma_{\Lambda}\cdot\sigma_N)\frac{1-P_r}{2}]e^{-(r/\beta_i)^2},
%\end{split}
\end{eqnarray}
where $P_r$ is the space exchange operator. The strengths $v_{0,even}^i$, $v_{\sigma\sigma,even}^i$,
$v_{0,odd}^i$ and $v_{\sigma\sigma,odd}^i$ are represented as functions of $k_F$ in Eq.(2.7) of Ref.\cite{1994YNG}.
The parameters of NF model are fitting to reproduce the observed binding energy of
$^5_\Lambda$He. However, it is pointed out that the NF model cause an overbound problem for the ground state of $^9_\Lambda$Be
due to the strong attraction of odd-state component of spin independent part of the $\Lambda$N interaction. In this case,
we tune the odd state part and $k_F$ to reproduce the observed binding energy of $^5_\Lambda$He and $^9_\Lambda$Be. The new parameters
of the modified NF model are listed in Table \ref{tab:YNG}.
\begin{table}[tbh]
\caption{$\Lambda$N interaction depth $v_{0,even}^i$ and $v_{0,odd}^i$
of the modified YNG-NF model. Here we take $k_F=0.963$ fm$^{-1}$.
}
\begin{tabular}{p{2cm}<{\centering}p{1.5cm}<{\centering}
p{1.5cm}<{\centering}p{1.5cm}<{\centering}}
  \hline\hline
 $\beta_i$(fm)& 1.50 &0.90 &0.50         \\
  \hline
 $v_{0,even}^i$ & -9.22 &-187.63&795.43  \\
 $v_{0,odd}^i$ & -5.67&-35.26&2141.79   \\
 \hline
\end{tabular}
\label{tab:YNG}
\end{table}

The Pauli principal between two $\alpha$ clusters is taken into account with the
orthogonality condition model(OCM). The OCM projection operator is represented by:
\begin{equation}
V_{\rm Pauli}= \lim\limits_{\lambda \rightarrow \infty}\lambda\sum_{f=0s,1s,0d}
|\phi_f ({\bf r}_{\alpha \alpha})><\phi_f ({\bf r'}_{\alpha \alpha})|.
\end{equation}
The Pauli forbidden states(0s,1s,0d) are ruled out when $\lambda$ is an infinity number
and practically the $\lambda$ is given around $\sim10^5$MeV, which is high enough to push
the unphysical states into a large energy region without affecting the physical states.

We use the $\alpha$-$\alpha$ interaction which reproduce the observed $\alpha$-$\alpha$ scattering
phase shift and the ground state of $^8$Be within the $\alpha$-$\alpha$ OCM. In this case, we fold
the modified Hasegawa-Nagata effective NN potential and pp coulomb potential into the $\alpha$ cluster
wave function.

\begin{figure}[htp]
\setlength{\abovecaptionskip}{0.cm}
\setlength{\belowcaptionskip}{-0.cm}
\centering
\subfigure{
%\label{fig:a}
\includegraphics[width=0.5\textwidth]{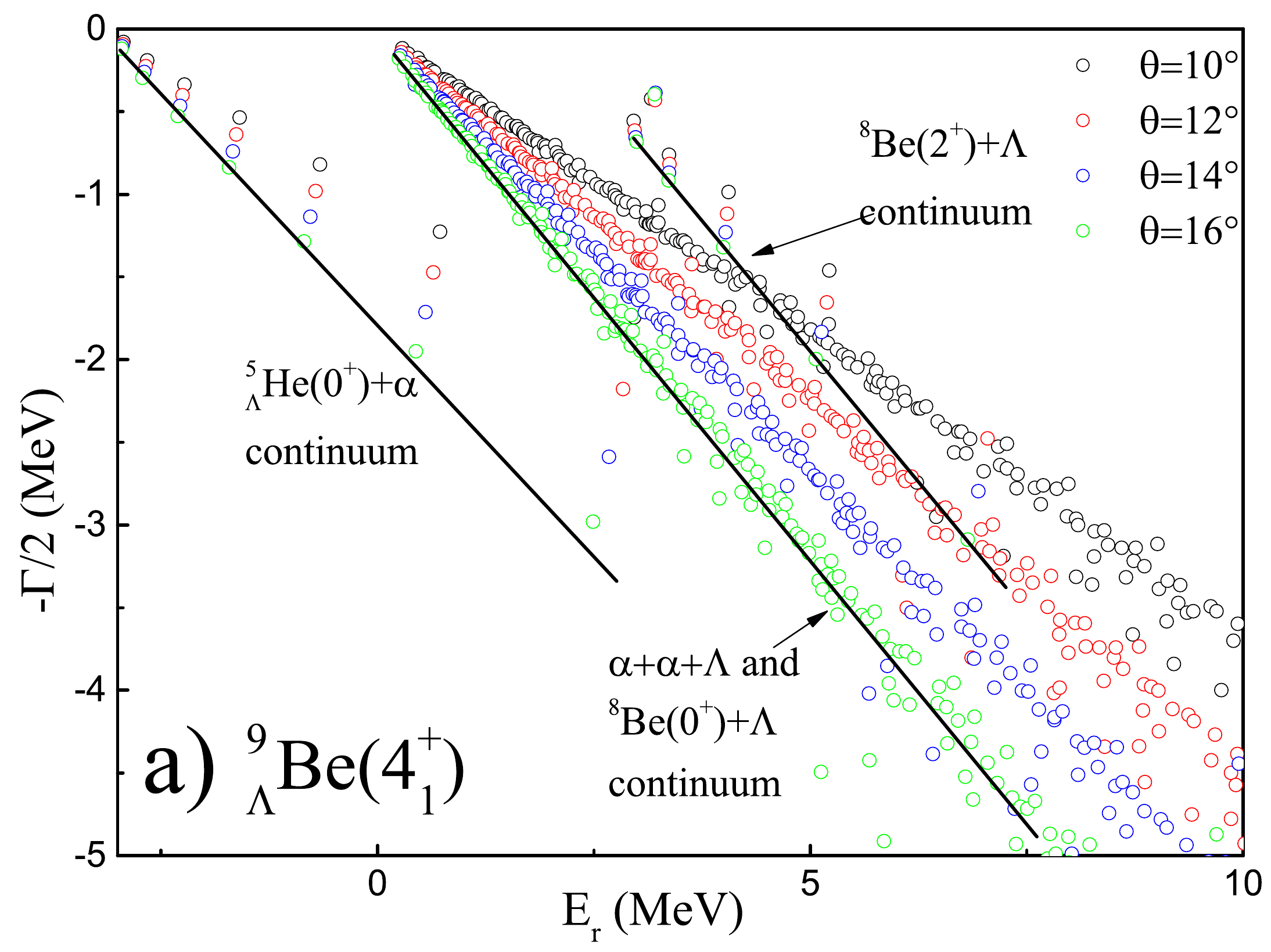}}
\hspace{1in}
\subfigure{
%\label{fig:b}
\includegraphics[width=0.5\textwidth]{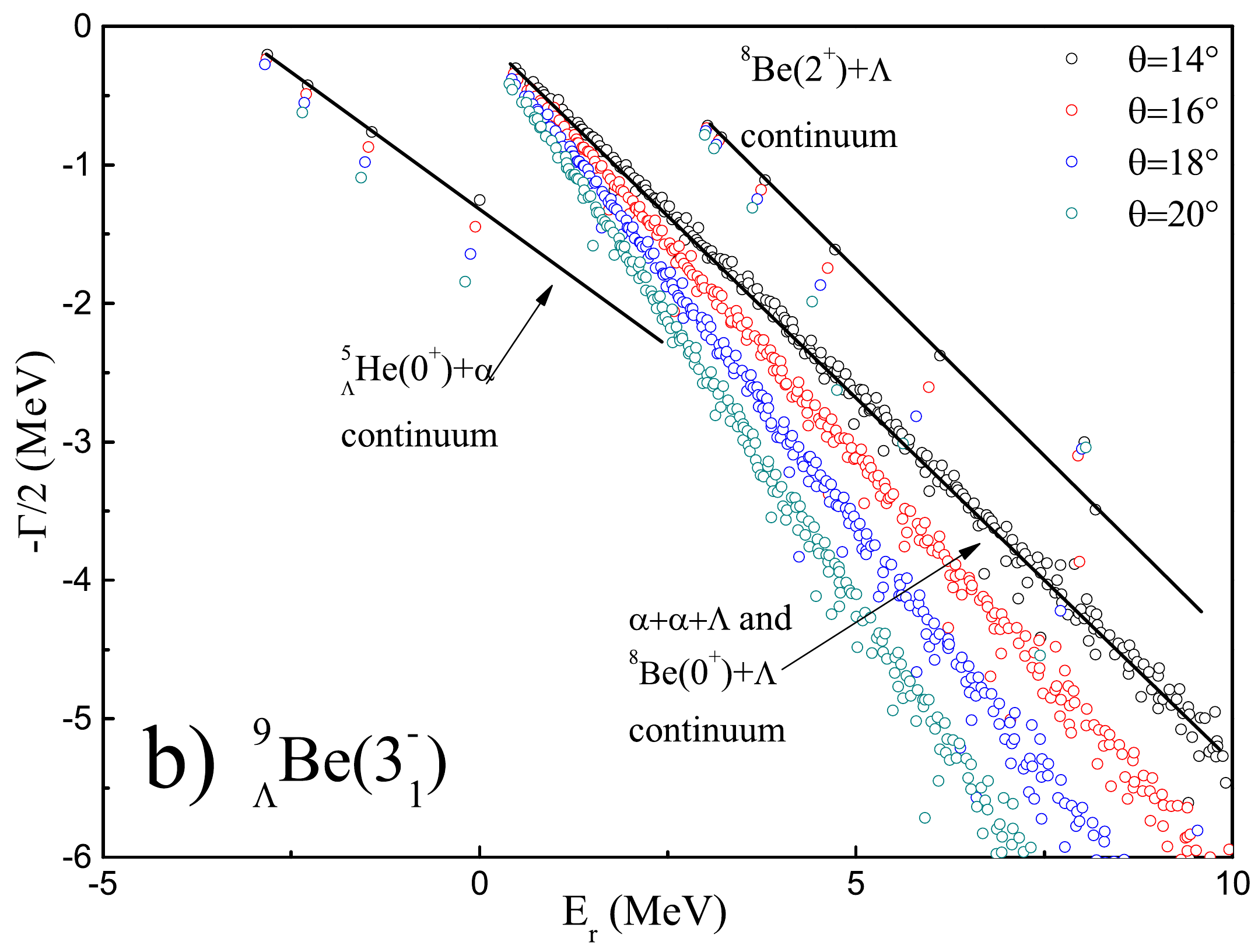}}
\caption{Dependence of the energy distribution on the
complex scaling angle $\theta$ for $^9_{\Lambda}$Be.
Two different cases are considered: (a) presence of a narrow resonance
with $J^{\pi}=4^+$ at $E_{r}=3.2$ MeV with $\Gamma=0.78$ MeV and (b)
presence of a broad resonance with $J^{\pi}=3^-$ at $E_{r}=9.4$ MeV with $\Gamma=7.1$ MeV }
\label{fig:csm} %% label for entire figure
\end{figure}

In this work, we calculate both bound state and resonant state and we employ a powerful method, named
complex scaling method(CSM) \cite{CSM1,CSM2,CSM3,CSM4,CSM5} with which we are capable of getting the energy, the decay width and the wave function
of a resonance. The CSM has been applied in nuclear physics for a long history \cite{csm-a,csm-b,csm-c,csm-d} and its validity has been proved
many times. By solving the complex scaled Schr$\ddot{\rm o}$dinger equation with a scaling angle $\theta$:
\begin{equation}
[H(\theta)-E(\theta)]\; \Psi(\theta)
\;= \; 0,
\end{equation}
where the scaling hamiltonian is obtained by setting:
\begin{equation}
r_c \rightarrow r_c e^{i \theta} , \quad  R_c \rightarrow R_c e^{i \theta},
\end{equation}
the energy and the width of the resonance is given as $E=E_r -i \Gamma/2$ which is independent of $\theta$. The bound state
will be stable in the negative real axis while the continuum states are rotated downwards at an angle of 2$\theta$ with the
real axis. In Fig.\ref{fig:csm}, we present two typical examples in calculating the resonant state of $^9_{\Lambda}$Be
using complex scaling method. In these two figures , the resonance remains stable when $\theta$ increases and the poles become
isolated from the continuum states. Moreover, we can see three lines for continuum states in Fig.\ref{fig:csm} which corresponds
to the $^5_\Lambda$He($0^+$)+$\alpha$ threshold, $\alpha$+$\alpha$+$\Lambda$ threshold or $^8$Be($0^+$)+$\Lambda$ threshold
which are close to each other and $^8$Be($2^+$)+$\Lambda$ threshold.

\section{Results and Discussion}
\subsection{Energy Spectra of $^9_\Lambda$Be}

\begin{figure*}[htp]
\setlength{\abovecaptionskip}{0.cm}
\setlength{\belowcaptionskip}{-0.cm}
\begin{center}
\centerline{\includegraphics[width=16.0 cm,height=10.0 cm]
                                        {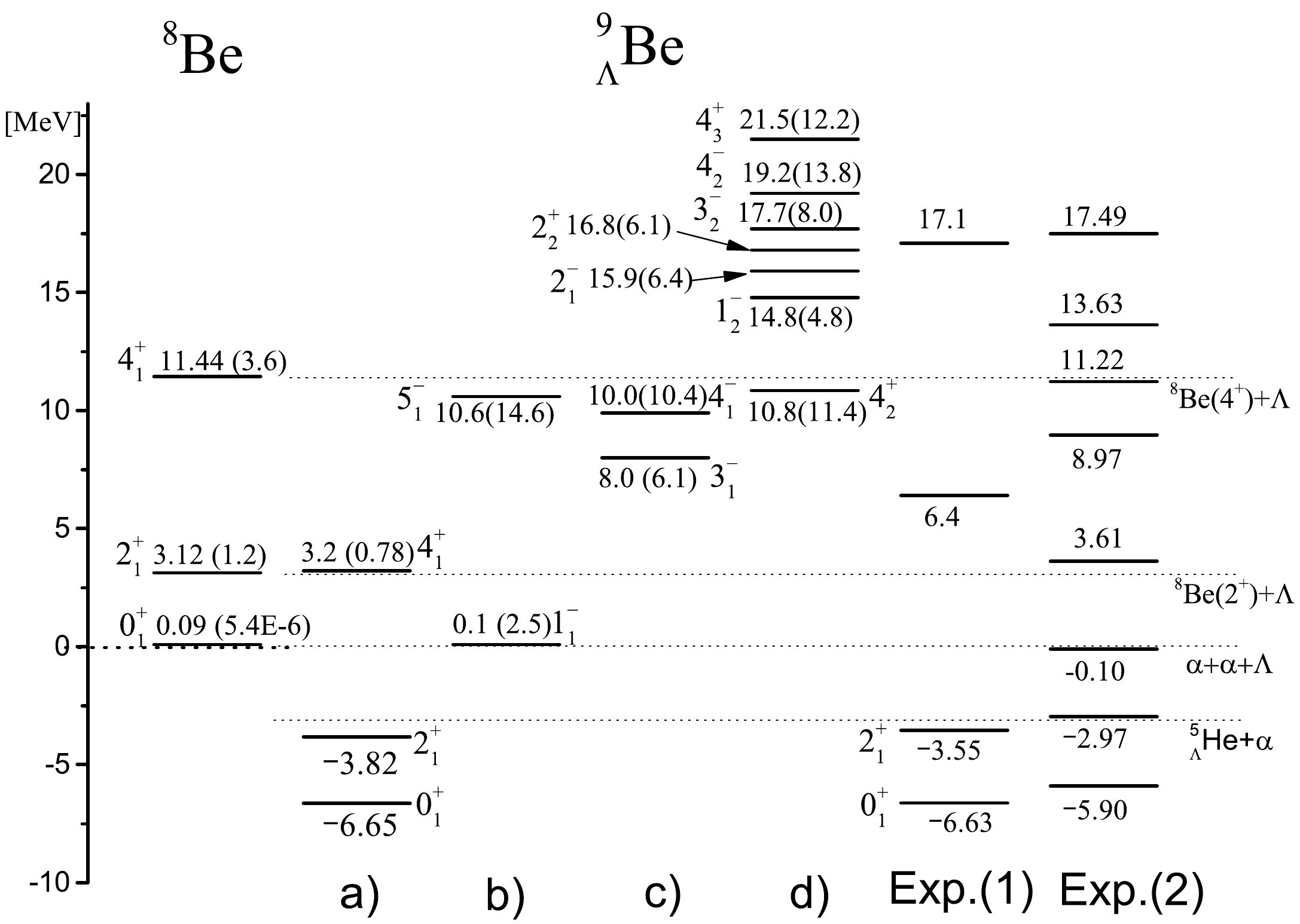}}
\end{center}
\caption{Calculated energy spectra of $^8$Be and $^9_{\Lambda}$Be with respect to $\alpha +\alpha+\Lambda$ three-body threshold. The values in parenthesis are decay widths. The spectra are categorized a) $^8$Be analogue band, b) genuine hypernuclear analogue, c) $^9$Be analogue and d) new analogue. The observed energies of $^9_\Lambda$Be in column Exp.(1) are taken from
Refs.\cite{be90p,be92p,be91m-exp}. And the observed energies of $^9_\Lambda$Be in column Exp.(2)
are taken from Refs.\cite{hashimoto-1998,Hhashimoto}.
}
\label{fig:spec}
\end{figure*}

The calculated energy spectra of $^9_{\Lambda}$Be are
shown in Fig.~\ref{fig:spec}. The left column shows
the resonant states, $0_1^+$, $2_1^+$ and $4_1^+$ states,
of $^8$Be, which are obtained by the CSM.
Next to the energy spectra of $^8$Be, we show the energy spectra of $^9$Be obtained by the OCM $+$ CSM. For later convenience, we group them into the column a), b), c), and d).
In all the subsequent calculations, the spin-orbit splitting of the $\Lambda$ particle and core is neglected since it is very small.

First we discuss the spectra of positive parity states of
$^9_{\Lambda}$Be, $0_1^+$, $2_1^+$ and $4_1^+$ states (column a).
We can see the $0_1^+$ and $2_1^+$ states are the bound states and are bound by $3.82$ MeV and $6.65$ MeV, respectively, relative to the $\alpha+\alpha+\Lambda$ threshold. Thus, the calculated $B_{\Lambda}=6.74$ MeV for the ground state and the excitation energy for the $2_1^+$ state, $E_{\rm ex}=2.83$ MeV, are in good agreement with the corresponding observed values, $B_{\Lambda}=6.71\pm0.04$MeV \cite{be90p} and $E_{\rm ex}=3.079\pm0.040$MeV \cite{be92p}, respectively. The $4_1^+$ state is obtained as a resonance, whose energy and width are calculated to be $E_r=3.2$MeV, above the $\alpha+\alpha+\Lambda$ threshold, and $\Gamma=0.78$MeV, respectively.
In Refs.\cite{motoba1985,bando1983,motoba1983}, the authors pointed out that these states are
$^8$Be analogue state, in which the $\Lambda$ particle couples in an $S$-wave to the $0_1^+$, $2_1^+$ and $4_1^+$ states of $^8$Be.
%In other words like SU(3) shell model configuration,  the analogue states correspond to $[s^5p^4](\lambda\mu)$=(40).

\begin{figure*}[htp]
\setlength{\abovecaptionskip}{0.cm}
\setlength{\belowcaptionskip}{-0.cm}
\begin{center}
\centerline{\includegraphics[width=16.0 cm,height=10.0 cm]
                                        {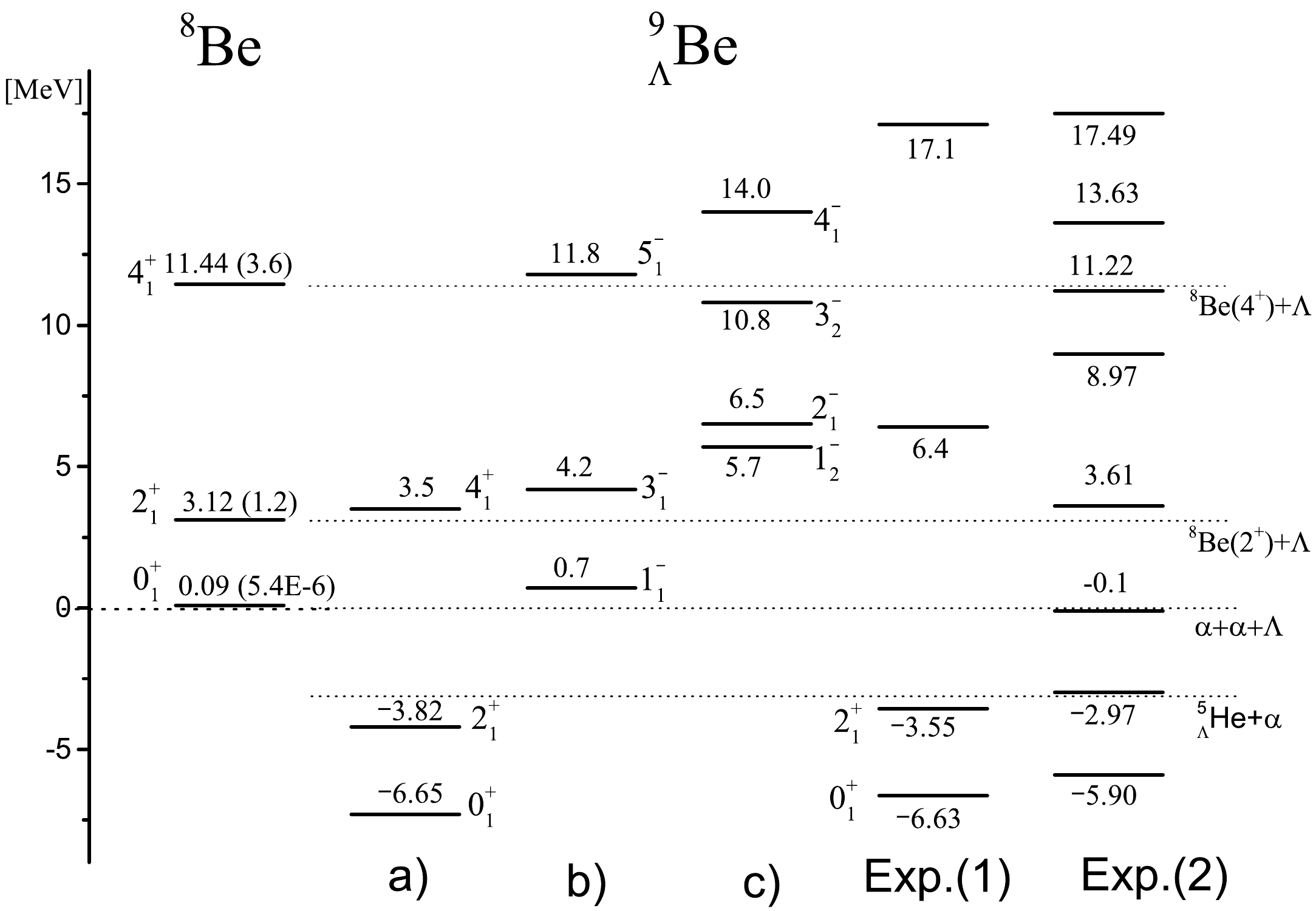}}
\end{center}
\caption{Calculate energy spectra of $^8$Be and $^9_{\Lambda}$Be by Motoba {\it et al.} \cite{motoba1985}. The spectra are categorized as a) $^8$Be analogue, b) genuine hypernuclear analogue, c) $^9$Be analogue.
The observed energies of $^9_\Lambda$Be in column Exp.(1) are taken from
Refs.\cite{be90p,be92p,be91m-exp}. And the observed energies of $^9_\Lambda$Be in column Exp.(2)
are taken from Refs.\cite{Hhashimoto,hashimoto-1998}.
}
\label{fig:spec-motoba}
\end{figure*}

In column b) of Fig.~\ref{fig:spec}, we show two resonant states, $1_1^-$  and $5_1^-$ state at
$E_r=0.1$ MeV with $\Gamma=2.5$ MeV and $E_r=10.6$ MeV with $\Gamma=14.6$ MeV, respectively.
These states may correspond to the so called 'genuine hypernuclear states' pointed out in Refs.~\cite{motoba1983,bando1983,motoba1985}, in which the $\Lambda$ particle occupies a $p$-orbit in a parallel direction to the $\alpha+\alpha$ axis of $^8$Be core. This $\Lambda$ particle motion is made possible because of no active effect of the Pauli principle of the $\Lambda$ particle to nucleons in the $^8$Be core, unlike the case of $^9$Be.
%In the the SU(3) shell model configurations, the $^9_\Lambda$Be genuine states are
%corresponding to $[s^4p^5](\lambda\mu)$=(50).

In column c) of Fig.~\ref{fig:spec}, we obtain the $3_1^-$ and $4_1^-$ resonant states at
$E_r=8.0$ MeV with $\Gamma=6.1$ MeV and $E_r=10.0$ MeV with $\Gamma=10.4$ MeV,
which can be categorized as `$^9$Be analog states'.
This means that in these states the $\Lambda$ particle occupies a $p$-orbit around the $^8$Be core, which is perpendicular to the $\alpha-\alpha$ axis, like a neutron orbiting around $^8$Be core in $^9$Be nucleus.

In column d) of Fig.~\ref{fig:spec},
we show several resonant states such as $4^+_2$, $1^-_2$, $2_1^-$, $2_2^+$, $3_2^-$, $4_2^-$ and $4_3^+$,
which are located at much higher energy region above the $\alpha + \alpha + \Lambda$ threshold. These states are not obtained in Refs.\cite{motoba1983,bando1983,motoba1985} and more details will be discussed later in subsection III.C.

%========================================
For comparison, in Fig.~\ref{fig:spec-motoba}, we show the energy spectra of $^9_{\Lambda}$Be obtained by Motoba {\it et al.}~\cite{motoba1985}, in which the YNG-NF potential for the $\Lambda$-nucleon interaction is adopted.
Here, we can see that their binding energy of the ground state is
overbound compared with the observed data. This overbinding is caused by a strong
attraction of odd-state part of the YNG-NF potential, as mentioned in Sec.II (see also Ref.\cite{hiyama1997ptp}).
We should note that their calculations are performed within the bound state approximation, in spite of the fact that almost all states shown here are located in an unbound region, with a finite decay width.

As pointed out by Band${\rm \bar{o}}$ and Motoba {\it et al.} \cite{motoba1983,bando1983,motoba1985},
the energy spectra of $^9_{\Lambda}$Be
are categorized in $^8$Be analogue states,
$^9_{\Lambda}$Be genuine states and $^9$Be analogue states,
which are shown in column a), b) and c) of Fig.~\ref{fig:spec-motoba}, respectively.
In column a), their binding energies of $0^+$, $2^+$ and $4^+$ states are similar to ours.
As for the $^9_{\Lambda}$Be genuine states in column b) of Fig.~\ref{fig:spec-motoba},
their energies of the $1_1^-$ and $5_1^-$ states are very close to the $1_1^-$ and $5_1^-$ states in our calculation shown in column b) of Fig.~\ref{fig:spec}.
We have to emphasize that in our calculation shown in Fig.~\ref{fig:spec},
we have no $3_1^-$ state corresponding to the $3^-$ state in $^9_{\Lambda}$Be genuine states in column b) of Fig.~\ref{fig:spec-motoba}.
In column c) of Fig.~\ref{fig:spec-motoba}, which can be categorized as the $^9$Be analogue states,
the binding energies of the $3_2^-$ and $4_1^-$ states are slightly higher than those of our $3_1^-$ and $4_1^-$ states shown in column c) of Fig.~\ref{fig:spec}, but both may correspond to each other.

On the other hand, we cannot find any $1^-$ and $2^-$ resonant states in the energy region of the $1_2^-$ and $2_1^-$ states shown in Fig.~\ref{fig:spec-motoba}.
The discrepancy of the energy spectra obtained by the resonance treatment like the present CSM and by the bound state approximation will be discussed in detail in the next subsection.

\subsection{$^8$Be analogue, $^9$Be analogue, and genuine states}

In this subsection, we discuss the structure of the states in columns a), b), and c), and also discuss
the reason why our spectra in columns b) and c) do not have one-to-one correspondence to those obtained by Band${\rm \bar{o}}$ and Motoba {\it et al.} in Refs.~\cite{motoba1983,bando1983,motoba1985}.

First, we find that the $0_1^+$, $2_1^+$ and $4_1^+$ states in column a) have the analogous structure to the $0_1^+$, $2_1^+$ and $4_1^+$ states of $^8{\rm Be}$, respectively, as is consistent with the results in Refs.~\cite{motoba1983,bando1983,motoba1985}. In fact, for these three states we calculate the $S$-wave components of the $\Lambda$ particle coupling to the $^8{\rm Be}$ core, which are found to be very large, $96\%$, $95\%$ and $93\%$, for the $0_1^+$, $2_1^+$ and $4_1^+$ states, respectively. This is nothing but the $^8{\rm Be}$ analogue structure, where the configurations $^8{\rm Be}(0^+)+\Lambda$, $^8{\rm Be}(2^+)+\Lambda$, and $^8{\rm Be}(4^+)+\Lambda$, are realized for the $0_1^+$, $2_1^+$, and $4_1^+$ states, respectively.
Band${\rm \bar{o}}$ and Motoba {\it et al.} also obtained in Refs.~\cite{motoba1983,bando1983,motoba1985} the similar values, around $95\%$, for these states, and hence we can say that our spectra in column a) well reproduce those in column a) shown in Fig.~\ref{fig:spec-motoba} obtained in Refs.~\cite{motoba1983,bando1983,motoba1985}.

As explained in Introduction, the classification of the spectra shown in Fig.~\ref{fig:spec-motoba} is, in a strong coupling limit, better understood by the nuclear SU(3) model, where the columns a), b), and c) correspond to the SU(3) irreducible representations, $(\lambda,\mu)=(4,0)$, $(5,0)$, and $(3,1)$, respectively. In order to investigate how much our spectra also keep this strong coupling SU(3)-like nature, we compare our wave functions  with the corresponding relative wave functions between the two-$\alpha$ clusters and $\Lambda$ particle described in terms of the SU(3) shell model picture.

From this aspect, we next discuss the $1_1^-$ state in column b) of Fig.~\ref{fig:spec}.
However, since this state is obtained by the CSM as having a very broad width, it is no more trivial to physically interpret any physical quantities calculated by using the resonant wave function. Thus we first construct an approximate $1_1^-$ wave function in a bound state region, so as to be smoothly connected to the resonant $1^-_1$ wave function with the broad width. That can be done by introducing the following attractive three-body force and artificially change the resonant wave function to a bound state wave function, to analyze the wave function without the difficulty,
\begin{equation}
V=V_0e^{-\mu(r_1^2+r_2^2+r_3^2)}, \label{eq:tbf}
\end{equation}
where $\mu$ is fixed to be $0.1$ fm$^{-2}$. When we choose $\emph{V}_0=-110$ MeV, the $1^-_1$ state becomes a weakly bound state, whose binding energy is $0.28$ MeV relative to the $^5_{\Lambda}$He+$\alpha$ threshold. We hereafter denote this artificial $1^-$ state as the $1_I^-$ state.

On the other hand, according to the Bayman-Bohr theorem \cite{Baymen1958}, the SU(3) $(\lambda,\mu)=(5,0)$ and $(\lambda,\mu)=(3,1)$ irreducible representations can be expressed below, in terms of the $\alpha$ cluster wave function,
%These representations can also be given in terms of the $\alpha$ cluster wave functions, as follows:
\begin{eqnarray}
&& |(0s)^4(0p)^4(0p)_\Lambda^1 (5,0)J=1 \rangle_{{\rm internal}} \nonumber \\
&& \hspace{3em} \propto \sum_{l=0,2}C_l^{(5,0)} \mathscr{A}|\, (4l,11)_{J=1} \phi_\alpha\phi_\alpha \rangle, \\
&& |(0s)^4(0p)^4(0p)_\Lambda^1 (3,1)J=1 \rangle_{{\rm internal}} \nonumber \\
&& \hspace{3em} \propto \sum_{l=0,2}C_l^{(3,1)} \mathscr{A}|\, (4l,11)_{J=1} \phi_\alpha\phi_\alpha \rangle,
\end{eqnarray}
where $C_l^{(\lambda,\mu)}=\langle (4,0)_l (1,0)_1|| (\lambda,\mu)_1 \rangle$, which are the reduced Clebsch-Gordan coefficients of SU(3) group for the vector coupling $(4,0)\otimes (1,0) = (5,0)\oplus (3,1)$, with $C_{l=0}^{(5,0)}=C_{l=2}^{(3,1)}=\sqrt{15/7}$ and $C_{l=2}^{(5,0)}=-C_{l=0}^{(3,1)}=\sqrt{15/8}$.
$\mathscr{A}$ is the antisymmetrization operator acting on the nucleons, $\phi_\alpha$ is the intrinsic wave function of the $\alpha$ particle, and $(nl,NL)_J$ is the harmonic oscillator wave functions for the relative motions between the two-$\alpha$ and $\Lambda$ particles defined below,
\begin{equation}
|(nl, NL)_J\rangle=[R_{nl}(\bm{r}_2), R_{NL}(\bm{R}_2)]_{J} \rangle. \label{eq:ovlp}
\end{equation}
Here the $\bm{r}_2$($\bm{R}_2$) are the Jacobian coordinate set of $C=2$ channel defined in Fig.~\ref{fig:be9job}.
We then define a normalized relative wave functions between the $\alpha$ clusters and $\Lambda$ particle, corresponding to the SU(3) irreducible representations,
\begin{eqnarray}
&& |\, (5,0)_1\rangle \equiv \sum_{l=0,2} C_l^{(5,0)}|\, (4l,11)_{J=1} \rangle, \nonumber \\
&& |\, (3,1)_1\rangle \equiv \sum_{l=0,2} C_l^{(3,1)}|\, (4l,11)_{J=1} \rangle, \label{eq:relwv}
\end{eqnarray}
and calculate the squared overlap between our $1_I^-$ state and the relative wave functions defined above, i.e. $|\, \langle (\lambda,\mu)_1|1_I^-  \rangle |^2$.
We obtain $0.45$ and $0.01$ for the $(5,0)_1$ and $(3,1)_1$ states, respectively, indicating that our $1_I^-$ state is much closer to the $(5,0)_1$ state than to the $(3,1)_1$ state (see Table~\ref{tab:1m}).

\begin{table}[htp]
\caption{Squared overlap of the `artificial' $1_I^-$ and $1_{II}^-$ states with the relative wave functions of the SU(3) shell model defined in Eq.~(\ref{eq:relwv}). See text for the definition of the `artificial' $1_{I}^-$ and $1_{II}^-$ states.}
\begin{tabular}{p{3cm}<{\centering}p{2cm}<{\centering}p{2cm}<{\centering}}
  \hline\hline
    $1^-_\lambda$  & $(5,0)_1$     & $(3,1)_1$    \\
  \hline
    $1_I^-$        & $0.45$        & $0.01$       \\
    $1_{II}^-$        & $0.01+0.001i$ & $0.39+0.02i$ \\
  \hline
\end{tabular}
\label{tab:1m}
\end{table}

%We then calculate the squared overlap of the approximate $1_1^-$ wave function with the relative wave functions given by the SU(3) $(5,0)$ and $(3,1)$ irreducible representations, corresponding to the genuine hypernuclear state and $^9{\rm Be}$ analog state, respectively, in a similar way to the case of the $^8{\rm Be}$ analogue states. \red{
%According to the Bayman-Bohr theorem \cite{Baymen1958}, the SU(3) irreducible representation corresponding to the $^8$Be analog states, $(\lambda,\mu)=(4,0)$ can be expressed below, in terms of the $\alpha$ cluster wave function,}
The reason why we obtain at most around $0.5$ for the squared overlap with the SU(3)-like configuration can be understood by comparing between the SU(3) $(\lambda,\mu)=(4,0)$ irreducible representation and our $0_1^+$, $2_1^+$, and $4_1^+$ wave functions.
In the same way to the $(\lambda,\mu)=(3,1),(5,0)$ cases, the SU(3) irreducible representation corresponding to the $^8$Be analog states, $(\lambda,\mu)=(4,0)$ can be given below, in terms of the $\alpha$ cluster wave function,
%These representations can also be given in terms of the $\alpha$ cluster wave functions, as follows:
\begin{equation}
|(0s)^4(0s)_\Lambda^1(0p)^4 (4,0)J \rangle_{{\rm internal}} \propto \mathscr{A}|\, (4J, 00)_J \phi_\alpha\phi_\alpha \rangle,
\end{equation}
 We then calculate the squared overlap between our wave functions for the $0_1^+$, $2_1^+$, and $4_1^+$ states and the harmonic oscillator wave functions for the relative motions between the $\alpha$ clusters and $\Lambda$ particle, $|\langle (4J,00)_{J}| J_1^+ \rangle |^2$.
We show in Table~\ref{tab:be8a} the squared overlap values, which are about $40\%$ for all the $0_1^+$, $2_1^+$ and $4_1^+$ states. Since in these states the $\Lambda$ particle couples to the $^8{\rm Be}$ core in an $S$-wave with almost $100$ \%, these rather small values indicate that the relative motion between the two-$\alpha$ clusters is excited strongly from the lowest $4\hbar \omega $ harmonic oscillator state, and the $\alpha$ clusters are loosely coupled to each other.
\begin{table}[htp]
\caption{Squared overlaps with the harmonic oscillator relative wave functions, of the $0_1^+$, $2_1^+$ and $4_1^+$.}
\begin{tabular}{p{4cm}<{\centering}p{3cm}<{\centering}}
  \hline\hline
 $J_1^+$ & $|\, \langle (4J,00)_J |J_1^+ \rangle |^2$ \\
  \hline
 $0_1^+$ & $37\%$    \\
 $2_1^+$ & $37\%$   \\
 $4_1^+$ & $39\%$    \\
 \hline
\end{tabular}
\label{tab:be8a}
\end{table}

Thus the squared overlap value $0.45$ is comparable to the values, around $0.40$, for the $0_1^+$, $2_1^+$, and $4_1^+$ states shown in Table~\ref{tab:be8a}.
We can therefore say that asymptotically the $1_1^-$ state shares a considerable fraction of the genuine hypernuclear structure and corresponds to the $1_1^-$ state in column b) in Fig.~\ref{fig:spec-motoba}.

Next let us compare in more detail our spectra with those obtained in Ref.~\cite{motoba1985}, shown in Fig.~\ref{fig:spec-motoba}. As mentioned in the previous section, we notice that in our spectra the $3^-$ state in column b), the $1^-$ and $2^-$ states in column c) are missing, to be one-to-one correspondent with the spectra in Fig.~\ref{fig:spec-motoba}.
It should, however, be noted that the spectra in Fig.~\ref{fig:spec-motoba} are obtained within the bound state approximation, while ours are obtained by taking into account the correct boundary condition of resonances. Then in order to understand how this discrepancy comes out, we demonstrate how the energy poles of the missing resonances disappear in the framework of CSM.

\begin{figure}[htp]
\setlength{\abovecaptionskip}{0.cm}
\setlength{\belowcaptionskip}{-0.cm}
\centering
\includegraphics[width=0.45\textwidth]{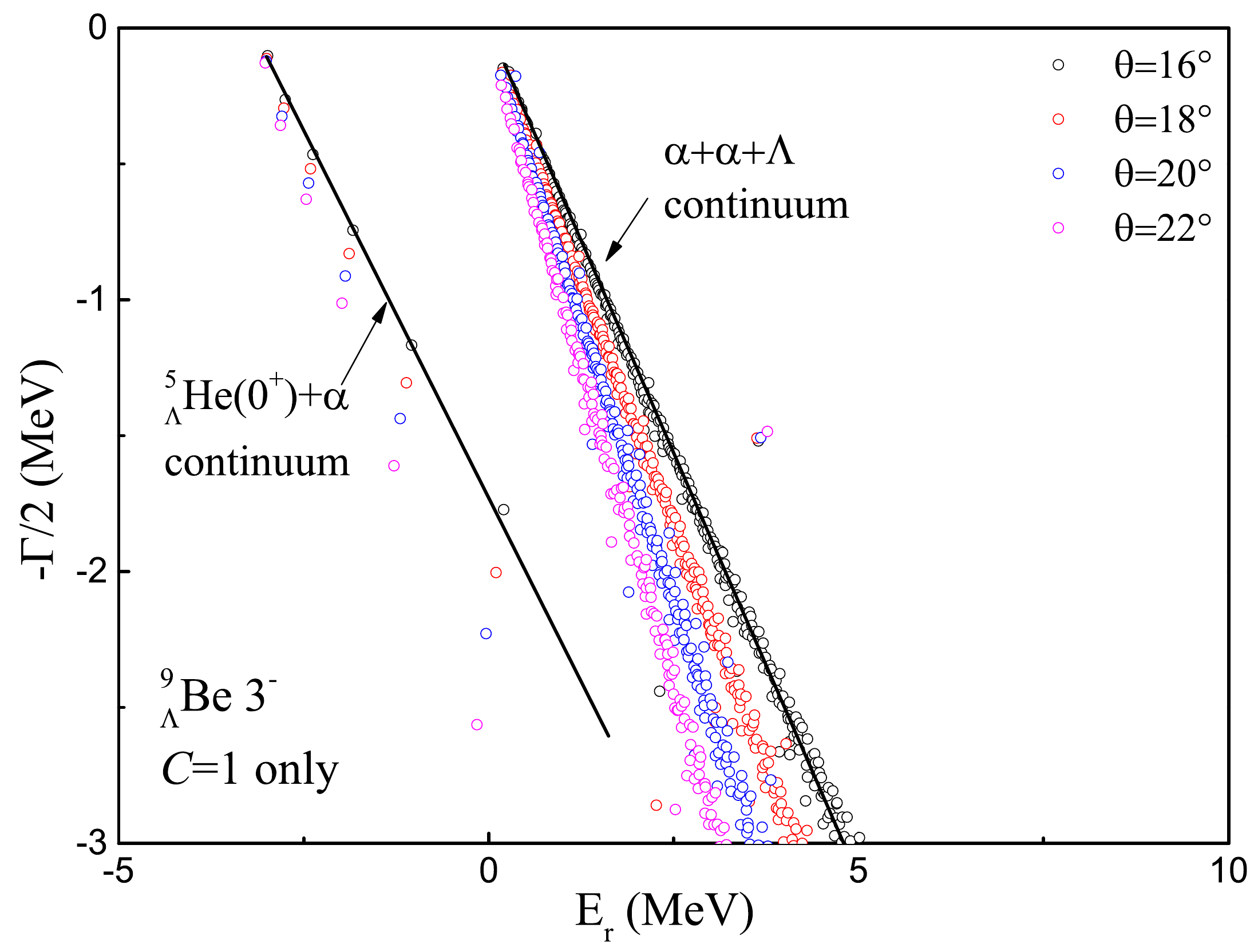}  %}
\caption{Dependence of the energy distribution on the
complex scaling angle $\theta$ for $^9_{\Lambda}$Be
in the case of $\emph{C}$=1 only. The $3^-$ state shows up
at $E_{r}=3.6$ MeV with $\Gamma=3.0$ MeV.
}
\label{fig:3MY} %% label for entire figure
\end{figure}
We first solve the present three-body problem in a restricted model space, where only the channel $C=1$ in Fig.~\ref{fig:be9job}, i.e. $^5_\Lambda{\rm He} + \alpha$ channel, is taken into account, to apply the CSM. The complex energies of $J^\pi=3^-$ states for several rotation angles $\theta$ are shown in Fig.~\ref{fig:3MY}. We can clearly see an energy pole at $E_r=3.6$ MeV with $\Gamma$=3.0MeV, which is similar to the energy of the $3_1^-$ state in column b) of Fig.\ref{fig:spec-motoba}. However, once we incorporate the $\emph{C}=2$ channel configurations in our three-body model space, the energy pole cannot be distinguished any more with the continuum states, and disappears. Since the inclusion of configurations of channel $C=2$ makes easier to incorporate $^8{\rm Be}-\Lambda$ configurations, it is considered that the ``missing'' $3^-$ state is dominated by a large amount of continuum components in $^8{\rm Be} + \Lambda$ channel, and it can no more survive as a resonance with a reasonable width. In fact, the calculated energy $E_r=3.6$ MeV is slightly higher than the $^8{\rm Be}(2^+) + \Lambda$ threshold, and therefore it is natural that the ``missing'' $3^-$ state has a sizable $^8{\rm Be}(2^+) + \Lambda$ continuum component.

In the same way as the $3^-$ state, we also calculated $1^-$ and $2^-$ states in the restricted configurations with $C=1$ channel only, to apply the CSM. Then we found one sharp resonant state with $J^\pi=1^-$ at $E_r=4.6$ MeV with $\Gamma=2.6$ MeV and also one sharp resonant state with $J^\pi=2^-$ at $E_r=5.6$ MeV with $\Gamma=2.9$ MeV.(See in blue colour levels of $1^-$ and $2^-$ in column c) of Fig.6).

In order to clarify the difference of the $1^-$ state obtained in this way from the $1_1^-$ state obtained at $E_r=0.1$ MeV, or to elucidate the SU(3)-like nature of this $1^-$ state, we calculate the squared overlap with the harmonic oscillator wave functions, defined by Eq.~(\ref{eq:relwv}), corresponding to the SU(3) $(\lambda,\mu)=(3,1)$ irreducible representation. First as we have done for the $1_1^-$ state at $E_r=0.1$ MeV, we introduce the attractive three-body force of Eq.~(\ref{eq:tbf}), with the same parameters as the case of the $1_1^-$ state, i.e. $\mu=0.1\ {\rm fm}^{-2}$ and $\emph{V}_0=-110$ MeV. With this three-body force, the $1^-$ state gains more binding energy and becomes a much sharper resonance, with $E_r=2.8$ MeV and $\Gamma=0.5$ MeV, which we denote $1_{II}^-$ state. We then calculate the squared overlap with the states $|(5,0)_1 \rangle$ and $|(3,1)_1 \rangle$ defined in Eq.~(\ref{eq:relwv}), corresponding to the $^9{\rm Be}$ analog in the SU(3) model interpretation. We obtain complex values $0.01+0.001i$ and $0.39+0.02i$ for the $(5,0)_1$ and $(3,1)_1$ states, respectively. Here in this case, the squared overlap is defined by $\langle \widetilde{1}_{II}^- | (\lambda,\mu)_1 \rangle \langle (\lambda,\mu)_1 |1_{II}^-  \rangle$, where the complex conjugate is not taken in the bra-state. For both values, the imaginary parts are much smaller than the real parts, due to the narrower width of the state, so that we can safely discuss the physical quantity as usual, by taking only the real parts (see Table~\ref{tab:1m}). The real part of the latter, $0.39$, is much larger than the one of the former, $0.01$, indicating that this state much more resembles the $(5,0)_1$ state than the $(3,1)_1$ state, unlike the case of the $1^-_I$ state discussed above. This value, $0.39$ is similar to $0.45$, the value of squared overlap between the $1_1^-$ state obtained with the same three-body force and the $(3,1)_1$ state in Eq.(\ref{eq:relwv}). Thus, we can conclude that this inherently ``missing'' $1^-$ state in our more precise calculations with the correct resonance boundary condition, is of the $^9{\rm Be}$ analogue nature and corresponds to the $1_2^-$ state in column c) of Fig.~\ref{fig:spec-motoba}.

%In order to investigate whether or not the missing $1^-$ state is a family of $^9$Be analogue states, we calculate the overlapping probability between $1^-$ state and the shell model configuration defined by Eq. Since this $1^-$ state is a broad resonant state, then it is expected that the wave function should have long-range tail. Therefore, as done in the case of $3^-$ state, by introducing three-body force in Eq.[13], we make the $1^-$ state compact. When we have $\emph{V}_0$=-110 MeV, missing $1^-$ state becomes a sharp resonant state at $E_r$=2.8MeV with $\Gamma$=0.5MeV. With this value, the overlapping probability with $|$31$>$ is about 40$\%$ which is significant. Then, we confirm that this $1^-$ state is the $^9$Be analogue state.

These $3^-$, $1^-$, and $2^-$ states are additionally shown in Fig.~\ref{fig:specY}, denoted in blue. Since it is difficult to analyze the resonance wave functions with broad widths, such as the $3_1^-$, $4_1^-$, and $5_1^-$ states, due to ill behaviour of their asymptotics, we just tentatively assign these states in column b) and c), i.e. the $5_1^-$ state in b) and $3_1^-$ and $4_1^-$ states in c). This assignment, however, seems to be reasonable since now we find the ``missing'' $3^-$, $1^-$, and $2^-$ states below the $3_1^-$, $4_1^-$, and $5_1^-$ states, which can compensate the missing spectra, to be consistent with the $J^\pi=1^-,\ 3^-,\ 5^-$ given by $(\lambda,\mu)=(5,0)$ and $J^\pi=1^-,\ 2^-,\ 3^-,\ 4^-$ given by $(3,1)$ SU(3) irreducible representations.

\subsection{New states of $^9_\Lambda$Be}

Besides the states displayed in column a)-c) of Fig.~\ref{fig:spec}, whose structures are studied by many authors, we newly find another three positive parity states, $2_2^+$, $4_2^+$ and $4_3^+$ states, and four negative parity states, $1_2^-$, $2_1^-$, $3_2^-$ and $4_2^-$ states, which are shown in columns d) and e) of the same figure. All these states are located at more than $10$ MeV above the $\alpha + \alpha + \Lambda$ threshold, with non-negligible widths as resonances, and have never been pointed out by the other authors before. We should note that these states would never be found without imposing a correct boundary condition of resonances, like the CSM in the present treatment of resonances.

\begin{figure*}[htb]
\setlength{\abovecaptionskip}{0.cm}
\setlength{\belowcaptionskip}{-0.cm}
\begin{center}
\centerline{\includegraphics[width=16.0 cm,height=10.0 cm]
                                        {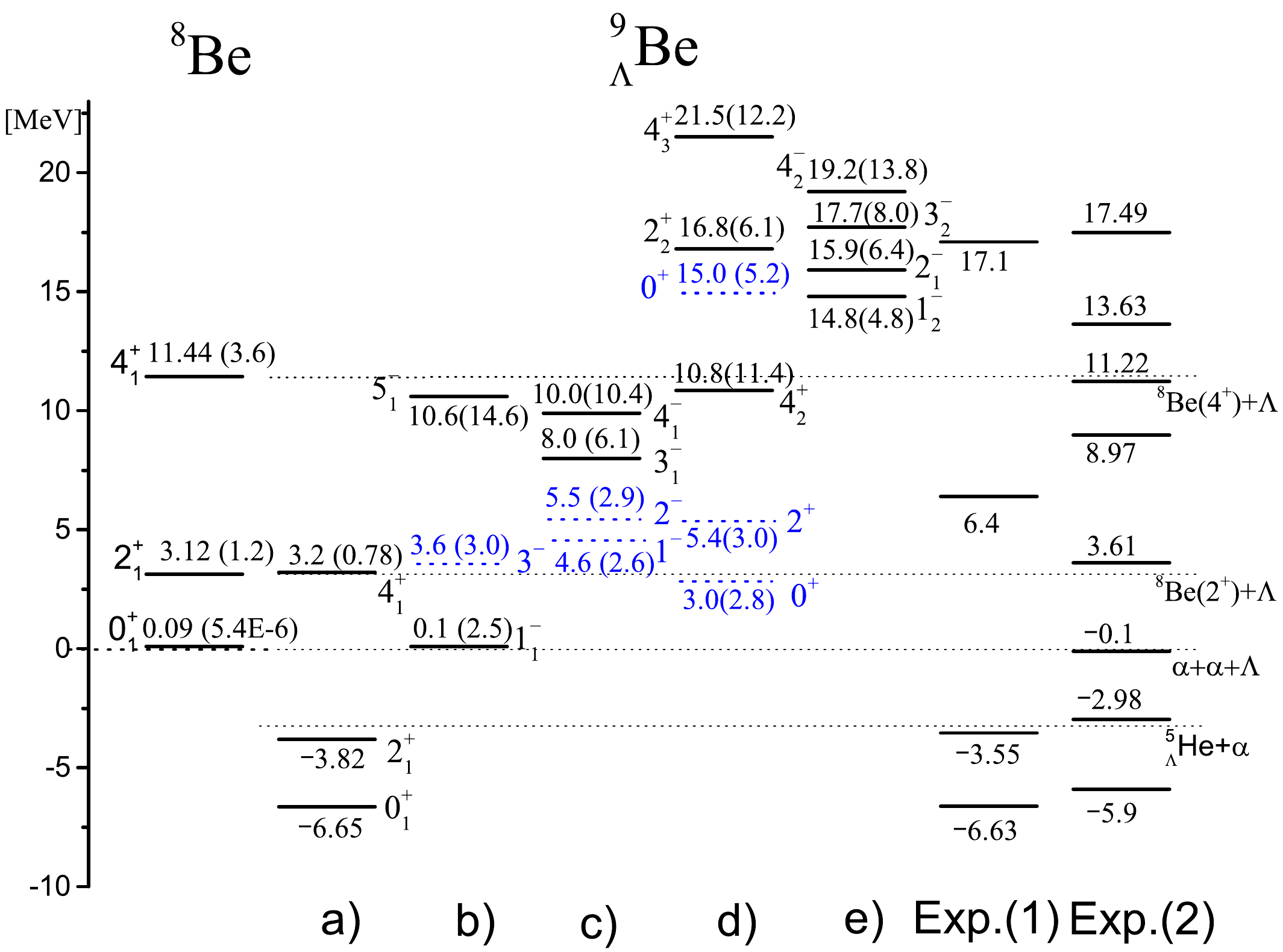}}
\end{center}
\caption{Calculated energy spectra of $^8$Be and $^9_{\Lambda}$Be with respect to
$\alpha +\alpha+\Lambda$ three-body threshold. The values in parenthesis are decay widths.
The spectra are categorized in a) $^8$Be analogue band, b) genuine hypernuclear analogue,
c) $^9$Be analogue, d) new states of positive parity and  e) new states of negative parity.
The states colored in black are calculated in both $\emph{C}$=1 and $\emph{C}$=2 channels while the
blued ones are calculated in only $\emph{C}$=1 channel. The observed energies of $^9_\Lambda$Be in column Exp.(1) are taken from
Refs.\cite{be90p,be92p,be91m-exp}. And the observed energies of $^9_\Lambda$Be in column Exp.(2)
are taken from Refs.\cite{Hhashimoto,hashimoto-1998}.}
\label{fig:specY}
\end{figure*}

As was mentioned in the previous subsection, these states also have broad widths and it is difficult to practically deal with the resonant wave functions. However, as was done in the previous subsection, for further information about the group of the new states, we solve the three-body problem with a practically restricted model space, with only $\emph{C}$=1 rearrangement channel, and search for further complex energy poles by the CSM. Then another three resonances show up, two $0^+$ and one $2^+$ states, at $3.0$ MeV with $\Gamma$=2.8MeV, $5.4$ MeV with $\Gamma=3.0$ MeV, and $15.0$ MeV with $\Gamma=5.2$ MeV, respectively. They are shown in column d) of Fig.~\ref{fig:specY} denoted by blue. The reason why the additional resonances appear when solved in the practically restricted model space may be similar to that of the case of the negative parity states discussed in the previous subsection. The inclusion of the channel $\emph{C}=2$ may increase the $^8{\rm Be}+\Lambda$ channel components, and eventually its continuum-like components as well, to make the states difficult to survive as resonances with a $^8{\rm Be}+\Lambda$-like structure. Thus, together with the artificial three states, as shown in column d), two groups of the $0^+$, $2^+$, and $4^+$ states may exist and each group seems to form a rotational band, possibly of $^8{\rm Be}+\Lambda$ structure.

%We calculate the following $S^2$-factor for these three states into $^{8}{\rm Be} + \Lambda$ channels.
%\begin{equation}
%{\cal S} = \sum_{L} \int|{\cal Y}_L(\hat{\bm{R}}_2)|^2R_2^2dR_2,
%\end{equation}
%with
%\begin{equation}
%{\cal Y}_L(R_2)=\left\langle\Psi_{J}(^9_{\Lambda}Be)\bigg{|} \left[
%\frac{\delta(R_2^\prime-R_2)}{R_2^{\prime 2}} Y_L(\hat{\bm{R}}_2^\prime), \varphi(^8\rm{Be})
%\right]_{J} \right\rangle,
%\end{equation}
%For the first $0^+$ state, we obtain a sizable value for the $^8$Be($0^+$)+$\Lambda$ channel, $0.36$. For the $2^+$ state, the sum of the $S^2$-factors into the $^8{\rm Be}(2^+)+\Lambda$ channels with $L=0,2,4$ is calculated to be $0.38$. For the second $0^+$ state, the sum of the $S^2$-factors into the $^8{\rm Be}(4^+)+\Lambda$ channels with $L=0,2,4, 6??$ is calculated to be $0.55$. Considering the fact that these states are obtained by solving the three-body problem without the  $\emph{C}=2$ rearrangement channel, i.e. $^8{\rm Be}+\Lambda$ channel, these sizable values indicate that the inclusion of the channel $\emph{C}=2$ may increase the $^8{\rm Be}+\Lambda$ channel components, and eventually its continuum-like components as well, to make the states difficult to survive as resonances with a $^8{\rm Be}+\Lambda$-like structure. Thus, together with the artificial three states, as shown in column d), two groups of the $0^+$, $2^+$, and $4^+$ states may exist and each group seems to form a rotational band, possibly of $^8{\rm Be}+\Lambda$ structure.

Finally, in Table.\ref{tab:exp2}, we compare our results with the
 observed data obtained by KEK with $(\pi^+,K^+)$ reaction in 1998 \cite{hashimoto-1998,Hhashimoto}. First, the observed binding energy of KEK(E336) is $B_{\Lambda}=5.99\pm 0.07$MeV, which is quite different from our results even we include the systematic error, $\pm0.36$MeV \cite{Hhashimoto}. In another way, this observed data is also quite different from another observed data in Ref.\cite{be90p}, $B_{\Lambda}=6.71\pm0.04$MeV, which is consistent with our results. We still don't know the reason for this difference and it remains further work. And for the first excited states, $E_{\rm ex}=2.93\pm0.07$MeV, is similar with our results, $E_{\rm ex}=2.83$MeV and another experimental data in Ref.\cite{be92p}, $E_{\rm ex}=3.079\pm0.07$MeV.

Second, for the resonances, our calculated genuine hypernuclear state, $1_1^-$ state is consistent with the 3rd experimental state. The observed 4th state, 3.61MeV, seems correspond to our $4^+_1$ state. The 5th observed data, 8.97MeV, is close to our $3^-_1$ , $^9$Be analog state. In addition, the observed 6th state, 11.2MeV, is close to our $4^+_2$ state and the 7th state, 13.63 MeV, corresponds to $1^-_2$. And the 8th observed data, 17.49MeV, is nearby  our $2^-_1$ state or $3^-_2$ state.
However, note that we do not calculate $(\pi^+,K^+)$ reaction.
To compare with our results with the experimental data,
it is necessary to calculate reaction cross section.
The calculation is our future work.

\begin{table}[htp]
\caption{Energy spectra of $^9_\Lambda$Be with respect to $\alpha-\alpha-\Lambda$ threshold. We present our calculated
energy together with the decay width for resonant states. The KEK(E336) are the experimental data obtained in 1998 by KEK \cite{hashimoto-1998,Hhashimoto}. All energies are given in MeV.}
\begin{tabular}{p{2cm}<{\centering}p{2cm}<{\centering}p{2cm}<{\centering}p{2cm}<{\centering}}
  \hline\hline
 $^9_\Lambda$Be &\multicolumn{2}{c}{Present work} & KEK(E336)  \\
 \hline
 &$E_r$&$\Gamma$&$E_r$ \\
 \hline
  $0^+_1$& $-6.65$ & -  & $-5.90\pm0.07$ \\
   $2^+_1$& $-3.82$ &- &$-2.97\pm0.07$  \\
  $1^-_1$&0.1& $2.5$  &$-0.10\pm0.13$ \\
  $4^+_1$& $3.2$ & $0.78$  &$3.61\pm0.13$ \\
  \hline
\end{tabular}
\label{tab:exp2}
\end{table}

\section{Summary}

We have calculated energy spectra of $^9_{\Lambda}$Be within the framework of
$\alpha +\alpha +\Lambda$ three-body model.
In this work, we employed the $\alpha-\alpha$ interaction
so as to reproduce the observed  $\alpha$$\alpha$ scattering data.
The Pauli
forbidden states(0s,1s,0d) between two $\alpha$s are ruled out by orthogonality condition model(OCM).
We employed the $\alpha$$\Lambda$ potential by folding procedure of YNG-NF $\Lambda N$ with $\alpha$ wave function.
Here, we adjust even- and odd-states of $\Lambda N$ interaction so as to reproduce the observed binding energies of ground states in
$^5_{\Lambda}$He and $^9_{\Lambda}$Be. For the resonant states of $^9_{\Lambda}$Be, we employed the CSM which is one of the powerful methods to obtain energy pole and decay width.

As a result, we categorize the level structure obtained here into (a) to (e):
(a) is $^8$Be-analogue states, (b) is genuine hypernuclear states, (c) is $^9$Be analogue states, which are pointed out by Band${\rm \bar{o}}$ and Motoba {\it et al.} \cite{motoba1983,bando1983,motoba1985},
and (d) and (e) are new states which have never been pointed out by
Band${\rm \bar{o}}$ and Motoba {\it et al.} \cite{motoba1983,bando1983,motoba1985}.
The points emphasized here are as follows:

1) The calculated binding energy of the $2^+_1$ state is $-3.82$ MeV, which does not contradict with the data, $-3.55$ MeV.
The calculated $4^+_1$ state is resonant state to be 3.2 MeV with $\Gamma=0.78$ MeV.
Here note that Motoba {\it et al.} \cite{motoba1985} obtain resonant energy only with bound state approximation.
Our calculated energies of $^8$Be analogue states, $0^+$, $2^+$, $4^+$ states are the same as those by Motoba {\it et al.}\cite{motoba1985}.
To confirm these three states are  $^8$Be analogue states,
we calculate $S$-wave components of the $\Lambda$ particle
coupling to the $^8$Be core and find the component to be
96 \%, 95\% and 93 \%, which are similar values by Motoba {\it et al.} \cite{motoba1985}.

2) The calculated first $1^-$ state is obtained by $0.1$ MeV above
$\alpha +\alpha +\Lambda$ three-body model with $\Gamma =2.5$ MeV.
To analyze the wave function of $1^-$ state,
by introducing three-body force to make bound state of this state,
we calculate the squared overlap of our wave function and
relative wave function by $(\lambda, \mu)=(5,0)$ and (3,1) represented by
SU(3). We find the value of (5,0) is 45 \% and value of (3,1) is 1\%.
Then, we confirm the $1^-_1$ is genuine hypernuclear state.

3) As shown in b) and c) of Fig.~6, the $3^-$, $1^-$ and $2^-$ states are missing which
are different from those by Motoba {\it et al.}  \cite{motoba1985}.
To analyze it, we solve three-body problem of $^9_{\Lambda}$Be
with a restricted model space, that is, with only $C=1$ channel in Fig.1
and we find resonance states. From the fact that inclusion of $C=2$ channel
causes disappearance of the resonant state, we find that
due to the large overlap of $^8{\rm Be}+\Lambda$
structure, the three resonant states are melted into continuum states.
Also, by analyzing the wave function of `missing' $1^-$ state,
we confirm that the states of c) of Fig.6 are categorized `$^9$Be analogue'
states which was pointed out by Band${\rm \bar{o}}$ and Motoba {\it et al.} \cite{motoba1983,bando1983,motoba1985}.

4) We obtain new states of positive parities and negative parities which have never been pointed out by Band${\rm \bar{o}}$ and Motoba {\it et al.} \cite{motoba1983,bando1983,motoba1985}  as shown in d) and e) of Fig.~6.
These states are located by around 10 MeV to 20 MeV with respect to $\alpha+\alpha +\Lambda$ three-body threshold with larger decay widths.

5) Finally, we compare the data with observed data so far obtained.
It is striking that our result of $1^-$ state, genuine hypernuclear state,
is consistent with the data with $0.1$ MeV,
The observed data of $E=3.61$ MeV is close to our energy of $4^+_1$ state
and observed $E=8.97$ MeV is close to calculated value of 8.0 MeV.
Our $5^-_1$, $4^-_1$ and $4^+_2$ states correspond to observed $E=11.22$ MeV
and observed $E=13.67$ MeV is close to $1^-_2$ state.
The observed $E=17.49$ MeV is nearby our  $2^+_2$ and $3^-_2$ states.
 To confirm observed states, it is necessary to calculate $(\pi^+,K^+)$ reaction. This calculation is our future work.

\begin{acknowledgments}
This work is supported by RIKEN-MOST collaboration
and Grants-in-Aid for
Scientific Research on Innovative Areas (No. JP18H05407,
No. JP18H05236 and No. JP18K03658) and JSPS Grant-in-Aid for Scientific Research(JP16H03995). One of the author(Q.W.) is supported by the National Natural Science Foundation of China (11475085, 11535005, 11690030) and National Major state Basic Research and Development of China (2016YFE0129300).
\end{acknowledgments}

\end{document}